\documentclass{aa}

\usepackage{graphics}

\def \hii{\ion{H}{ii}}
\def \hi{\ion{H}{i}}
\def \hei{\ion{He}{i}}
\def \di{\ion{D}{i}}
\def \nii{[\ion{N}{ii}]}
\def \oii{[\ion{O}{ii}]}
\def \oiii{[\ion{O}{iii}]}
\def \oi{\ion{O}{i}}

\def \ha{H$\alpha$}

\def \hg{H$\gamma$}

\def \he{H$\epsilon$}
\def \hz{H$\zeta$}
\def \da{D$\alpha$}
\def \db{D$\beta$}
\def \dg{D$\gamma$}

\def \de{D$\epsilon$}
\def \dz{D$\zeta$}
\def \dt{D$\eta$}
\def \kms{${\rm km}\,{\rm s}^{-1}$}

\def \eg{{\it e.g.}}

\def \cm2{cm$^2$}

\begin{document}

   \thesaurus{08                
             (02.12.2           
              09.01.2           
              09.08.1           
              09.09.1 M42       
              09.09.1 M8        
              09.09.1 M16       
              09.09.1 M20       
              09.09.1 M17       
              09.09.1 DEM S 103)}

\headnote{Letter to the Editor \\
Manuscript intended for the December 1, 2000 Special Edition \\
(Early Science with the VLT: The opening of Kueyen)}

   \title{Revealing deuterium Balmer lines in \ion{H}{ii} 
regions with VLT-UVES~\thanks{Based on observations 
collected at the European Southern 
Observatory, Paranal, Chile [ESO VLT-UT2 N$^o$~65.I-0498(A)].}
}

  \author{      G.~H\'ebrard \inst{1}
          \and
                D.~P\'equignot \inst{2}
          \and
                J.~R.~Walsh \inst{3}
          \and
                A.~Vidal-Madjar \inst{1}
          \and
                R.~Ferlet \inst{1}
          }

   \offprints{Guillaume H\'ebrard}

   \institute{Institut d'Astrophysique de Paris, CNRS,
              98 bis Boulevard Arago, F-75014 Paris, France 
              (hebrard@iap.fr, vidalmadjar, ferlet).
         \and
              Laboratoire d'Astrophysique Extragalactique et de 
              Cosmologie associ\'e au CNRS (UMR 8631) et \`a l'Universit\'e 
              Paris 7, DAEC, Observatoire de Paris-Meudon, F-92195 
              Meudon C\'edex, France (daniel.pequignot@obspm.fr).
         \and
              Space Telescope European Co-ordinating Facility, 
              European Southern Observatory, Karl-Schwarzschild-Strasse 2,
              D-85748 Garching bei M\"unchen, Germany 
              (jwalsh@eso.org).
             }

   \date{Received ? / Accepted ?}

   \maketitle

   \begin{abstract}

The search for deuterium Balmer lines with VLT-UVES
is reported in \hii\ regions of the Galaxy and the Magellanic Clouds. 
The \di\ lines appear as faint, narrow emission features in the 
blue wings of the \hi\ Balmer lines and can be distinguished 
from high-velocity \hi\ emission. 
The previous identification to deuterium is re-inforced beyond doubt. 

The detection of \da\ and \db\ in Orion 
(H\'ebrard et al.~\cite{hebrard00}) is confirmed and 
deuterium lines are now detected up to at least \dt. 
The UVES observations provide the first detection of 
Balmer \di\ lines in four new \hii\ regions 
(M~8, M~16, M~20, and DEM~S~103 in~SMC), 
demonstrating that these lines are of common occurence.

   \end{abstract}

%

\section{Introduction}

Deuterium is an element of primordial origin. Measuring 
its abundance in different astrophysical sites brings 
valuable constraints on the Big-Bang nucleosynthesis 
and the Galactic evolution 
(\eg\ Lemoine et al.~\cite{lemoine99}).

The detection and identification of the deuterium Balmer lines 
\da\ and \db\ in emission in the Orion Nebula was first 
reported by H\'ebrard et al.~(\cite{hebrard00}, hereafter Paper~I). 
The narrowness of these lines, their strength with respect to 
the hydrogen lines and finally their relative fluxes 
were incompatible with recombination excitation, but 
could be understood in terms of fluorescence excitation by 
stellar UV continuum in the Photon Dominated Region (PDR), 
located behind the ionized region.

Here, observations of the whole Balmer series with 
the new spectrograph UVES, installed at the Nasmyth focus of VLT-UT2, 
are presented for Orion and other \hii\ regions.
Observations are described in Sect.~\ref{observations}. Results 
for each \hii\ region are presented in Sect.~\ref{results}. New evidence 
in support to the identification of deuterium is discussed in 
Sect.~\ref{discussion}. A more complete analysis 
will follow in forthcoming papers.

\section{Observations}
\label{observations}

Observations were secured during the night 2000 July 25th-26th,   
using the UV-Visual Echelle Spectrograph (UVES) located at the 
Nasmyth focus of Kueyen, the second VLT Unit Telescope 
(D'Odorico \& Kaper~\cite{uves_manual}). 
Spectra from both the red and blue arms were registered simultaneously on 
two detectors, using the standard setting DIC1~(390+564). The approximate 
spectral ranges were 3290\AA~-~4530\AA\ (blue arm), and 4610\AA~-~5620~\AA\ 
and 5660\AA~-~6660\AA\ (red arm), encompassing the whole Balmer series. 

The slits were 
8\arcsec\ and 11\arcsec\ long for the blue and red arms respectively.
The slit width was 1\arcsec\ on the sky. 
According to the staff of the VLT, 
the spectral resolution was $R=\lambda/\Delta\lambda\simeq40\,000$ 
(Full Width at Half Maximum, FWHM), equivalent to $\sim7$\kms. 
The present conclusions do not depend on the exact value of $R$, 
which will be determined after reducing the calibration exposures. 
A total exposure time of one hour was devoted to each \hii\ region 
(except for Orion, Sect.~\ref{m42}), the 
observations being divided in short sub-exposures to prevent 
detector saturation at the \hi\ Balmer lines.

Data reduction (bias subtraction, flat-fielding, 
wavelength calibration) was performed with the UVES pipeline, 
using the available calibration database.   
1D spectra were box-extracted from the central third of the slits. 
The standard sky-subtraction algorithm, 
inappropriate for extended objects, was omitted. 
This extraction was judged robust enough 
for this preliminary study. 
Subsequent data reduction will be performed over the whole slit length, 
using the calibration exposures obtained during the observing run. 
 
Cosmics and bad pixels were cleaned where necessary. 
Sub-exposures were averaged (no shift was observed 
from one sub-exposure to the next) and the 
lines were shifted to the same radial velocity. 
For a given object, the peak fluxes of the different lines 
were assigned the same value in order to display the relative 
variations of the weak lines (figures of Sect.~\ref{results}). 
Shifts and normalizations were all based on 
Gaussian fits to the emission lines.

\section{Results}
\label{results}

Line detections reported here in the blue wings of the \hi\ lines 
are at least at the 5-$\sigma$ confidence level. 
Most of them are confirmed by the detection of lines at 
the same velocity for several principal quantum numbers $n$.

\subsection{Orion Nebula (\object{M 42})}
\label{m42}

The area observed in Orion 
was the same as the one observed previously (Paper~I). 
The slit, oriented North-South, was located 2.5\arcmin\ South of 
$\theta^1$~Ori~C (\object{HD 37022}) at coordinates 
$\alpha=$05:35:16.7, $\delta=-$05:25:29 (J2000).
The exposure time was 30~min in the red arm and 50~min in the blue arm.

\begin{figure}[ht]
\resizebox{\hsize}{!}{\includegraphics{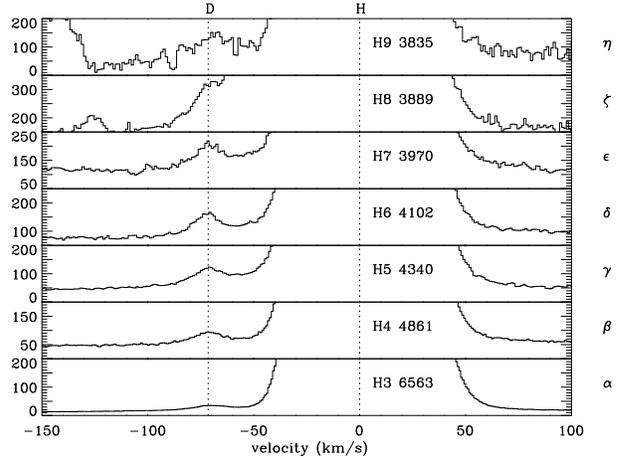}}
\caption[]{
Wings of \ha\ to H$\eta$ (noted H3 to H9) 
in the Orion Nebula. All \hi\ lines 
are centred at 0\kms\ velocity (right dotted line) and are normalized 
to identical peak intensities ($2.1\times10^4$ on $y$-scale). 
The dotted line to the left corresponds to the wavelengths 
adopted for the \di\ lines (Table~\ref{lines}). 
\hz\ is blended with \hei. 
}
\label{m42_fig}
\end{figure}

Plots of the \hi\ Balmer line wings are shown in Fig.~\ref{m42_fig}. 
Deuterium lines are detected from \da\ to \dt. They are redshifted 
$\sim10$\kms\ with respect to \hi\ (the isotopic shift between 
\di\ and \hi\ at rest is $-81.6$\kms), 
in good agreement with the previous measurements (Paper~I). 
\di\ lines seem to be detected up to D16, but elaborate treatment 
is required due to low signal-to-noise ratio.

FWHM are from Gaussian fits, after quadratic subtraction of the 
instrumental point-spread function. 
The FWHM of the \di\ lines is $\sim11$\kms, much 
less than that of the \hi\ recombination lines ($\sim30$\kms). 
Widths similar to these were found for the 
lines detected in M~8, M~16, M~20 and DEM~S~103 (see below). 
From Fig.~\ref{m42_fig}, it is apparent that \di\ increases relative 
to \hi\ for increasing $n$, at least up to \de. 
Approximate relative fluxes are given in Table~\ref{table_m42}. 
Despite lower signal-to-noise ratio, a similar trend 
exists in the data for M~8 and M~16. 

\begin{table}
\caption[]{Preliminary line flux ratios in M~42}
\label{table_m42}
\begin{tabular}{cc|cc|cc}
\hline
line &    \di/\hi\      & line &      \di/\hi\    & line &  
    \di/\hi\       \\
\hline
 $\alpha$ & $2\times10^{-4}$ &  $\gamma$ 
& $7\times10^{-4}$ & $\epsilon$  &  $10\times10^{-4}$\\
 $\beta$ & $6\times10^{-4}$ &  $\delta$ & $9\times10^{-4}$ &  &  \\
\hline
\end{tabular}
\end{table}

\begin{figure}[ht]
\resizebox{\hsize}{!}{\includegraphics{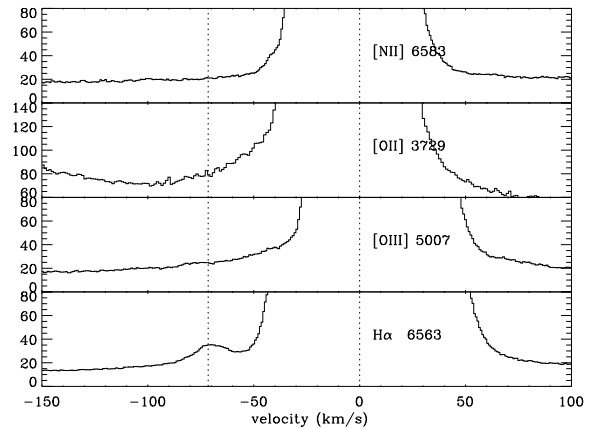}}
\caption[]{Same as Fig.~\ref{m42_fig} for wings of 
\nii, \oii, \oiii\ and \ha\ in Orion
(peak intensities 2.1$\times$10$^4$).
Compare to Fig.~\ref{m17_other_fig}.
}
\label{m42_other_fig}
\end{figure}

In Fig.~\ref{m42_other_fig}, are shown on the same scale the wings 
of \nii\ $\lambda$6583\AA, \oii\ $\lambda$3729\AA, 
\oiii\ $\lambda$5007\AA\ and \ha. No counterparts to \di\ can be seen 
for lines other than \ha, thus excluding any interpretation 
in terms of emission from high-velocity ionized gas 
(see Sect.~\ref{m17}). 
Similarly, counterparts are lacking 
in M~8, M~16, M~20 and DEM~S~103. 

\subsection{Lagoon Nebula (M 8)}

\begin{figure}[ht]
\resizebox{\hsize}{!}{\includegraphics{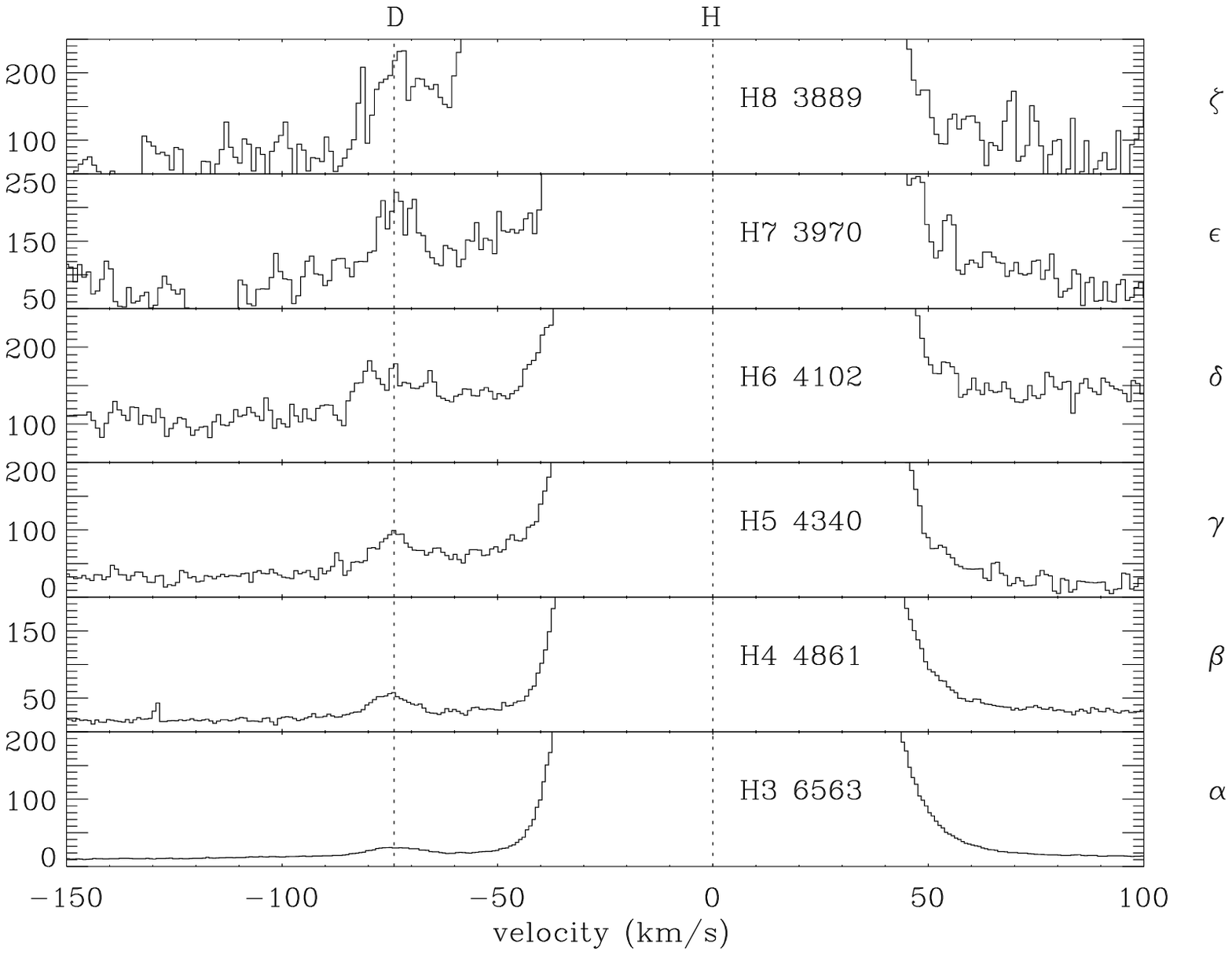}}
\caption[]{
Same as Fig.~\ref{m42_fig} for wings of 
\ha\ to H$\zeta$ in M~8, 
but with peak intensities 2.0$\times$10$^4$. 
}
\label{m8_fig}
\end{figure}

In \object{M 8}, the slit was 
oriented North-South and located 17\arcsec\ East and 18\arcsec\ 
North of Herschel~36 (\object{HD 164740}), at 
$\alpha=$18:03:40.8, $\delta=-$24:22:25. This position 
corresponds to position L11 in Bohuski~(\cite{bohuski73}). 
Deuterium is detected from \da\ to \dz\ (Fig.~\ref{m8_fig}).
Again the flux ratios range from 
$F({\rm D}\alpha)$/$F({\rm H}\alpha)\simeq2\times10^{-4}$ to 
$F({\rm D}\zeta)$/$F({\rm H}\zeta)\simeq1\times10^{-3}$. 

M~8 was observed at a second slit position: 43\arcsec\ South from 
Herschel~36 [position L7 in Bohuski~(\cite{bohuski73}), not shown here]. 
The coordinates were $\alpha=$18:03:40.3, $\delta=-$24:23:27 and 
the slit was oriented East-West.
Only \da\ was detected, with a weaker 
flux: $F({\rm D}\alpha)$/$F({\rm H}\alpha)\simeq3\times10^{-5}$.

\subsection{Eagle Nebula (M 16)}

\begin{figure}[ht]
\resizebox{\hsize}{!}{\includegraphics{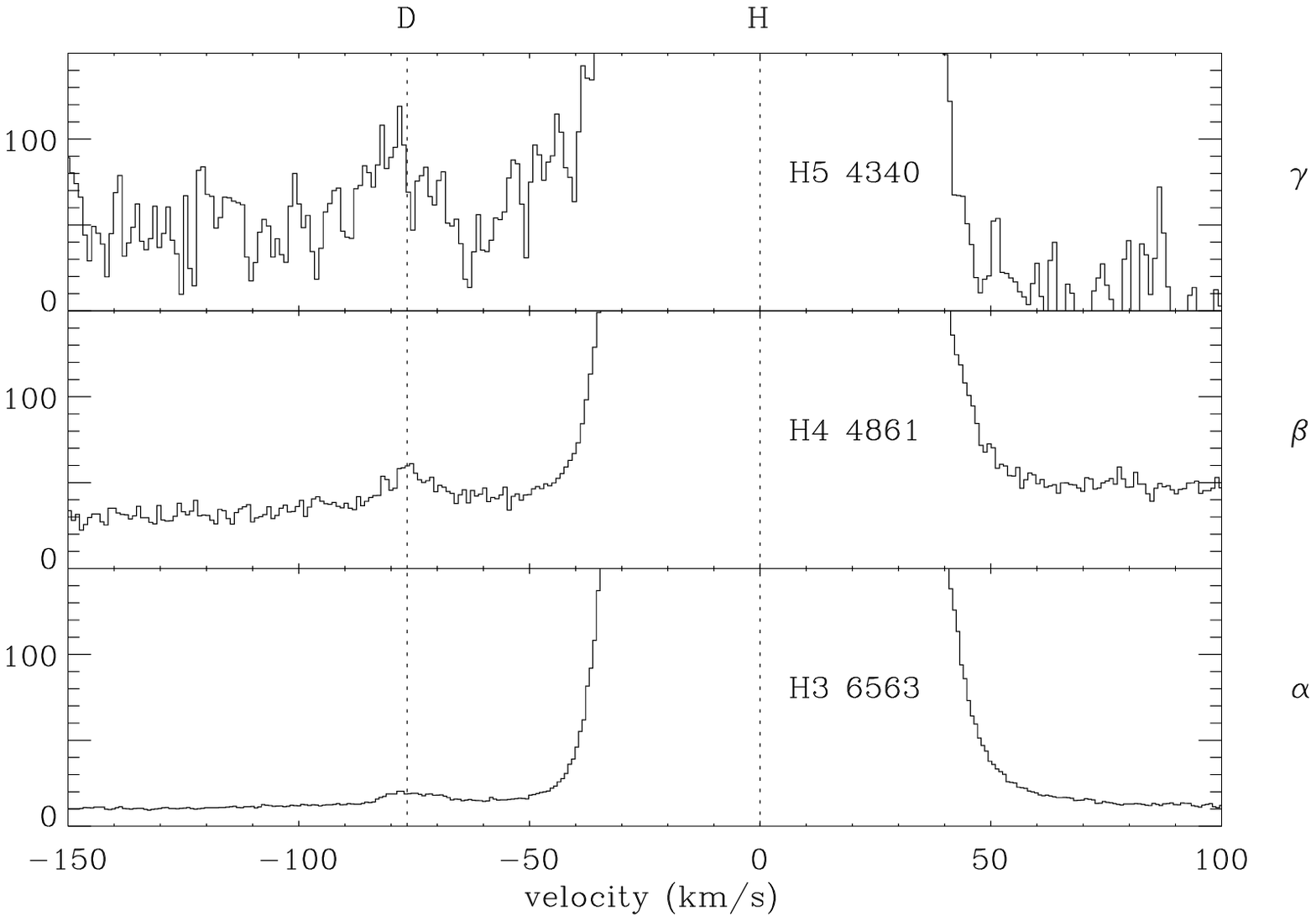}}
\caption[]{Same as Fig.~\ref{m42_fig} for wings of 
\ha\ to \hg\ in M~16, 
but with peak intensities $1.1\times10^4$. 
}
\label{m16_fig}
\end{figure}

In \object{M 16}, the slit was oriented 
North-South and located at $\alpha=$18:18:51.7, $\delta=-$13:49:07. 
It corresponds to one of the brightest regions of the PDR in this nebula 
(Levenson et al.~\cite{levenson00}).
Deuterium is detected from \da\ to \dg\ (Fig.~\ref{m16_fig}).
The flux ratios range from 
$F({\rm D}\alpha)$/$F({\rm H}\alpha)\simeq2\times10^{-4}$ to 
$F({\rm D}\gamma)$/$F({\rm H}\gamma)\simeq1\times10^{-3}$. 

\subsection{Trifid Nebula (M 20)}

\begin{figure}[ht]
\resizebox{\hsize}{!}{\includegraphics{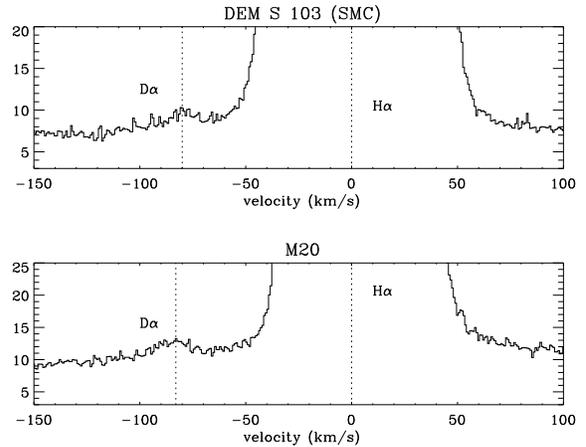}}
\caption[]{
Same as Fig.~\ref{m42_fig} for wings of 
\ha\ in M~20 (bottom, peak intensity 4000)
and DEM~S~103 in \object{SMC} (up, peak intensity 3500). 
}
\label{m20_n66_fig}
\end{figure}

In \object{M 20}, the slit was 
oriented North-South and located 51\arcsec\ East and 
23\arcsec\ South from \object{HD 164492}, at 
$\alpha=$18:02:27.3, $\delta=-$23:02:14.
This position corresponds to position T12 in Bohuski~(\cite{bohuski73}). 
Only \da\ was detected (Fig.~\ref{m20_n66_fig}).

\subsection{DEM~S~103 in the Small Magellanic Cloud}

The last deuterium Balmer line detection was performed outside 
the Galaxy, in the 
brightest \hii\ region of the SMC, namely 
\object{DEM S 103} [Henize~66, Caplan et al.~(\cite{caplan96})]. 
The coordinates of the slit, oriented North-South, were 
$\alpha=$00:58:51.6, $\delta=-$72:10:09. 
Again, only \da\ was detected (Fig.~\ref{m20_n66_fig}).

\subsection{Omega Nebula (M~17): high-velocity structure emission}
\label{m17}

\begin{figure}[ht]
\resizebox{\hsize}{!}{\includegraphics{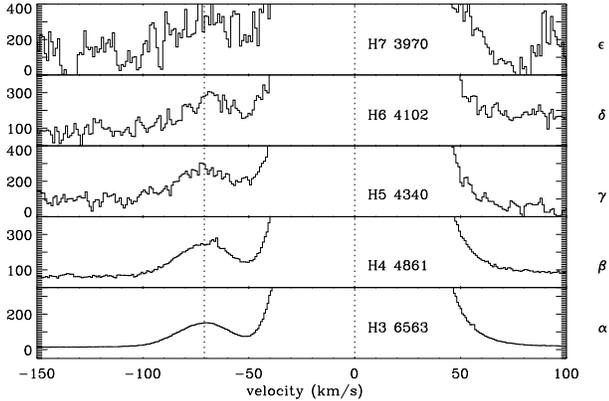}}
\caption[]{
Same as Fig.~\ref{m42_fig} for wings of 
\ha\ to \he\ in M~17 (peak intensities $2.1\times10^4$). 
Here the blue-shifted features are not identified with \di\ but 
with \hi\ emission from a high-velocity ionized structure. 
}
\label{m17_fig}
\end{figure}

\begin{figure}[ht]
\resizebox{\hsize}{!}{\includegraphics{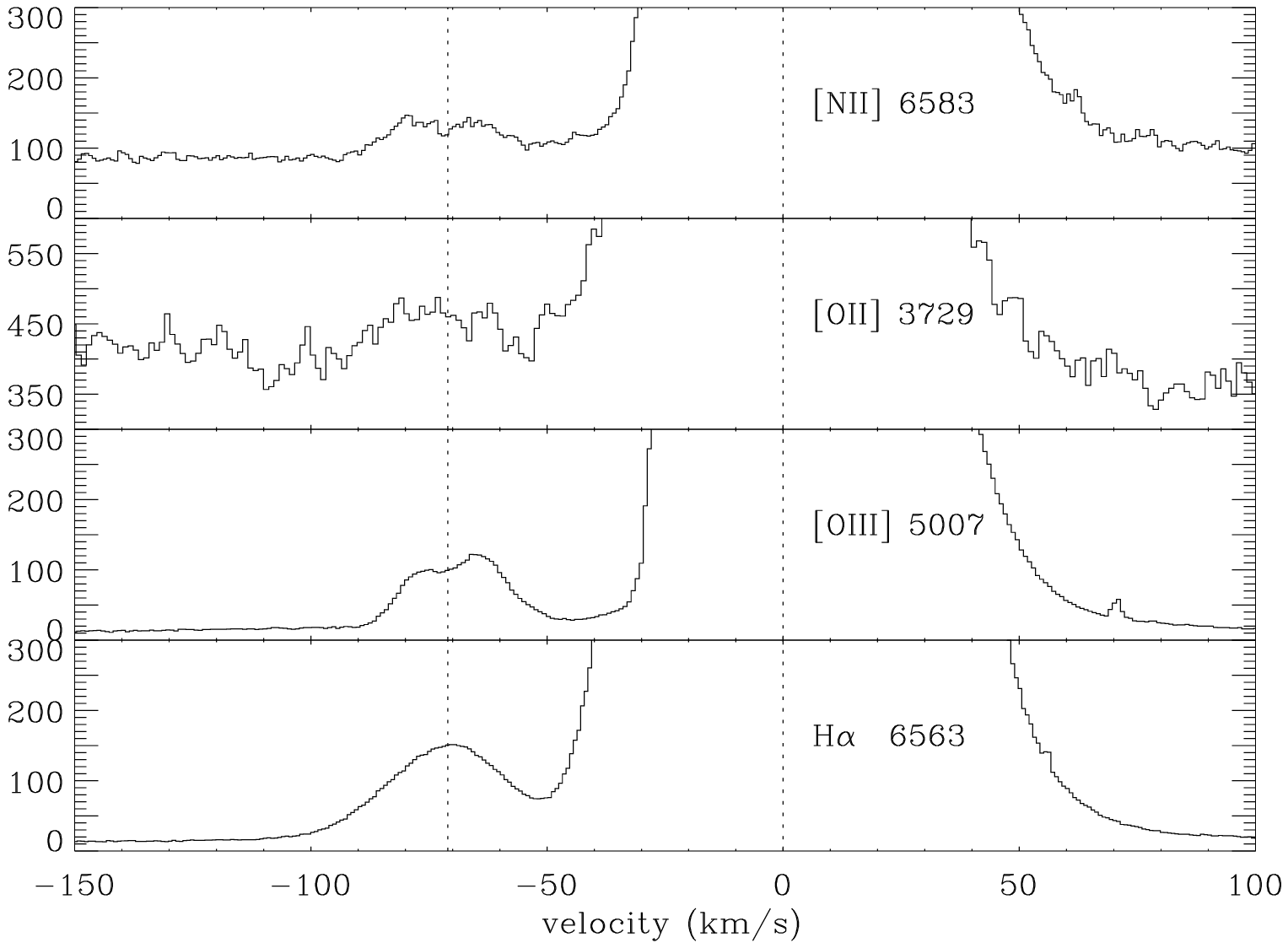}}
\caption[]{
Same as Fig.~\ref{m42_fig} for wings of 
\nii, \oii, \oiii\ and \ha\ in M~17 
(peak intensities $2.1\times10^4$). Here, the 
blue-shifted features have counterparts in ions. 
Compare to Fig.~\ref{m42_other_fig}.
}
\label{m17_other_fig}
\end{figure}

In \object{M 17}, the slit was oriented 
North-South and located at $\alpha=$18:20:48.0, $\delta=-$16:10:31. 
Here the emission features detected in the blue wings of the \hi\ lines, 
from \ha\ to \he\ (Fig.~\ref{m17_fig}), differ from those 
shown in previous targets: 
\begin{itemize}
\item
they are broad (FWHM $\simeq$ 20\kms, instead of $\sim$10\kms,
whilst the main \hi\ component has the usual FWHM $\simeq$ 30\kms);
\item
they are proportional to the \hi\ lines 
(intensity $\sim3\times10^{-3}$ relative to nearby \hi\ 
for every $n$); 
\item
\nii, \oii\ and \oiii\ present clear counterparts at the same velocity 
(Fig.~\ref{m17_other_fig}).
\end{itemize}

It is concluded that, in this case, the features should be mainly 
due to \hi\ emission from ionized material with velocity $\sim-70$\kms\ 
relative to the main body of the nebula. 
The width of these features is compatible 
with recombination excitation in a hot H$^+$ gas. 
In fact, Meaburn~\& Walsh~(\cite{meaburn81}) 
detected velocity components shifted by $\sim-70$\kms\ from the 
main \ha\ component, $\sim2$\arcmin\ South of the UVES position. 
Clayton et al.~(\cite{clayton85}) showed that the
high-velocity material was probably associated with a breakout
(collimated outflow) of a radially expanding shell.
A \da\ line with the same relative flux as in Orion would be 
one order of magnitude weaker than the feature near \ha, 
making detection difficult without an elaborate treatment. 

\section{Confirming the identification of deuterium}
\label{discussion}

In addition to the objects presented in Sect.~\ref{results}, two more 
\hii\ regions were observed: \object{30 Doradus} (in the LMC) 
and \object{Sh2-100}. In the former, the velocity field 
appears complex and more work is needed. In the latter, no obvious 
\di\ lines are seen, suggesting that no PDR is present along 
the line of sight. So far, \di\ lines have therefore been detected 
in at least five out of the eight \hii\ regions observed. 
Kinematic properties appear in Table~\ref{lines}. 

\begin{table}
\caption[]{Kinematic properties of the \di\ lines$^a$}
\label{lines}
\begin{tabular}{lcc|lcc}
\hline
Object    & Shift$^b$ & Width$^c$ & Object  & Shift$^b$ & Width$^c$ \\
\hline
M~42      & 10  & 11  & M~20      & -1: & 13: \\
M~8       &  8  &  9  & S~103     &  2: & 10: \\
M~16      &  5  &  9  & M~17      &  -  &  -  \\ 
\hline
\end{tabular}

$^a$ Colons indicate low-accuracy values (from \ha\ and \da\ only).\\
$^b$ Average shift of \di\ with respect to \hi\ (\kms).\\
$^c$ Average FWHM of the \di\ lines (\kms).\\
\end{table}

The new observations bring confirmatory evidence in favour of the 
identification of the deuterium Balmer lines. 
They complement and add to the results presented in Paper~I. 
The lines are seen with similar characteristics 
in the five \hii\ regions, using different telescopes (CFHT and VLT) 
and different spectrographs (GECKO and UVES). Instrumental artifacts 
(such as grating ghosts or in-order scattered light) can 
be definitely ruled out. 

Since the lines are seen for many members of the Balmer series, 
they can only be \di\ or blue-shifted \hi\ emission. \hi\ emission 
may arise from H$^+$ gas (recombination) or H$^0$ gas (fluorescence). 
High-velocity ionized structures will produce \hi\ recombination lines 
with properties like those already listed in the case of M~17 
(width, flux, counterparts), 
not observed in the other \hii\ regions described in Sect.~\ref{results}. 
A high-velocity neutral structure cannot be formally excluded for 
any one isolated object, but the probability that such a structure 
could exist and yet be detectable only in \hi\ is low. 
No evidence for the existence of such a structure could be 
found in the case of Orion (Paper~I). 
Considering the present data, it would be extraordinary if 
such a neutral component could be present in such a systematic manner 
in five different \hii\ regions, always at about the same velocity. 

Understandably, the \di\ lines are narrow since they 
arise from a cold material with small thermal velocity. 
Nonetheless, considering the prevalence of large velocity 
fields in \hii\ regions, it was not a priori obvious that these 
lines would appear so systematically narrow (Table~\ref{lines}). 
The explanation partly lies in the fact that 
the entrance aperture of UVES is relatively small
and that observable \hii\ regions tend to be 
incomplete on one side, with the associated molecular cloud 
and PDR located behind the expanding H$^+$ region. 
This is consistent with the tendency shown by the \di\ lines 
to be redshifted with respect to the \hi\ lines (Table~\ref{lines}). 
Thus, in practice, 
a small line width (at the expected wavelength!) 
turns out to be an important criterion to identify \di. 
On the other hand, \hii\ regions may exist with 
PDR's encompassing a large velocity range. A fundamental 
criterion for \di\ identification remains the lack of counterparts 
in lines from ionized species. 
Large variations of the line intensity ratio \di/\hi\ with $n$ 
constitute another useful criterion (Table~\ref{table_m42}), 
since fluorescence will 
generally not result in the same decrement as the one 
corresponding to recombination. 

As a result of the present high spectral resolution and high 
signal-to-noise observations, the identification of deuterium Balmer 
lines is now very safe. 

\section{Conclusion}

Detection of deuterium Balmer emission in five \hii\ regions 
is reported. These are first detections in four targets, 
including an extragalactic one. Detection 
was made feasible thanks to the large collecting area of 
the 8.2m VLT mirror and the high efficiency of UVES.
Fluorescence is confirmed as the probable excitation mechanism 
of \di, recombination being excluded. 
Spectroscopic criteria leading to virtually certain 
identification of \di\ in any given \hii\ region are 
now clearly established. 

Possible ways to determine D/H from \di\ Balmer lines 
were discussed in Paper~I. 
One method requires a knowledge of the $n$ for which 
the line ratio \di/\hi\ starts decreasing. The 
detection of \di\ up to D9, and possibly D16,  
in Orion suggests this as a promising way of investigation.
Comparison of \di\ to \oi\ fluorescence lines, present in the
UVES spectra and also produced in 
the PDR, may be another way to explore.

\begin{acknowledgements}
We thank the staff of the VLT for excellent assistance.
\end{acknowledgements}

\end{document}